\documentclass[submission,copyright,creativecommons]{eptcs}
\providecommand{\event}{FROM 2024} % Name of the event you are submitting to

\usepackage{iftex}

\ifpdf
  \usepackage{underscore}         % Only needed if you use pdflatex.
  \usepackage[T1]{fontenc}        % Recommended with pdflatex
\else
  \usepackage{breakurl}           % Not needed if you use pdflatex only.
\fi

\usepackage{macros}
\usepackage{macros-math}
\usepackage{macros-logic}
\usepackage{macros-IL}
\usepackage[T1]{fontenc}
\usepackage[utf8]{inputenc}
\usepackage{amssymb}
\usepackage{listings}
\usepackage{xpatch}
\usepackage{amsmath}

\usepackage{color}
\definecolor{keywordcolor}{rgb}{0.7, 0.1, 0.1}   % red
\definecolor{tacticcolor}{rgb}{0.0, 0.1, 0.6}    % blue
\definecolor{commentcolor}{rgb}{0.4, 0.4, 0.4}   % grey
\definecolor{symbolcolor}{rgb}{0.0, 0.1, 0.6}    % blue
\definecolor{sortcolor}{rgb}{0.1, 0.5, 0.1}      % green
\definecolor{attributecolor}{rgb}{0.7, 0.1, 0.1} % red

\xpretocmd\lstlisting{\vspace{-0.8\baselineskip}}{}{}

\title{Intuitionistic Propositional Logic in Lean}
\author{Dafina Trufaș
\institute{LOS, Faculty of Mathematics and Computer Science, University of Bucharest}
\institute{Institute for Logic and Data Science, Bucharest}
\email{dafina.trufas@s.unibuc.ro}
}
\def\titlerunning{Intuitionistic Propositional Logic in Lean}
\def\authorrunning{Dafina Trufaș}

\newenvironment{theorem}{\bgroup\par\noindent\textbf{Theorem.} }{\egroup}
\begin{document}
\maketitle

\begin{abstract}
In this paper we present a formalization of Intuitionistic Propositional Logic
in the Lean proof assistant. Our approach focuses on verifying two completeness
proofs for the studied logical system, as well as exploring the relation between the
two analyzed semantical paradigms - Kripke and algebraic. In addition, we prove a large number of theorems
and derived deduction rules.
\end{abstract}

\section{Introduction} 
We formalize Intuitionistic Propositional Logic (IPL) using the Lean interactive theorem prover \cite{deMoura}.
Our main goal is verifying the soundness and the strong completeness of IPL,
with respect to both the Kripke and the Heyting algebras semantics. The language we work with has falsity, conjunction, disjunction and implication
as primitive connectives and for syntactical inference we use the Hilbert-style proof system
introduced by Gödel in \cite{godel}.

For the formalization we present in this paper, we chose the Lean proof assistant \cite{deMoura}.
An evidence of Lean's proving power and versatility is the Mathlib library \cite{mathlib}, maintained by the Lean community.
This work aligns with the effort of the Mathlib community to encode mathematical knowledge, and particularly logical systems, in Lean.
The underlying theory of Lean is based on a version of dependent type theory, known
as the calculus of inductive constructions \cite{coquand}.
Thus, type-checking is the mechanism which assists the user in their approach
to prove mathematical statements, either by directly constructing proof terms or by using
Lean's so-called tactic-mode.

In the following, we describe the main stages of the implementation and
motivate our main design choices. Sections \ref{lang} and \ref{prf} describe the formalization of
the language and proof-system of IPL. The Kripke completeness proof is based on the
so-called canonical model, whose construction relies on the notion of disjunctive theory. Some results
about consistent and complete pairs, presented in Section~\ref{disj}, are also essential in the flow of this first completeness theorem. 
In the upcoming Section~\ref{kripke-sem}, we introduce the Kripke semantics, then
in Section~\ref{kripke-sound-compl} we present the main steps of the completeness formalized proof with respect to it.
Similarly, Section~\ref{alg} proceeds by defining the necessary Heyting algebras notions, establishes the algebraic semantics and concludes by proving the second completeness theorem and
establishing the equivalence between the validity notions.
Our presentation is inspired by the textbooks of Mints \cite{mints}, Fitting \cite{fitting} and Troelstra \cite{troelstra},
and the lecture notes of Kuznetsov\cite{kuznetsov} and Georgescu \cite{georgesculogic,georgescualgebras}.
All the detailed proofs can be found in my Bachelor's thesis, which is available online at 
\cite{trufas-thesis}.

To the best of our knowledge, the only proof of completeness for IPL
formally-verified in Lean is due to Guo, Chen and Bentzen \cite{bentzen}. However, the novelty of our approach consists in:
\be
\item using a different Hilbert-style proof system;
\item proving a large collection of theorems and derived deduction rules;
\item formalizing the algebraic semantics of IPL and proving a second completeness theorem, with respect to it;
\item implementing a semantic proof of the equivalence between algebraic and Kripke validity;
\item the manner we dealt with the countability of the set of formulas, which we consider simpler than the method in \cite{bentzen}.
\ee
\section{On the formalization}

The Lean code is structured in 8 files, which we briefly describe in the following.
First, we have the $Formula.lean$ file, which contains the definition of the language (Section~\ref{lang}),
as well as the proof of the countability of the $Formula$ type (Section~\ref{disj}). Then, the $Syntax.lean$ file
proceeds by formalizing the definition of $Proof$ (Section~\ref{prf}). It includes a large collection of theorems and derived deduction rules,
as well as the deduction theorem and some utilitary lemmas.
The $Semantics.lean$ file contains the definition of the Kripke model, and the semantical definitions we detail in Section~\ref{kripke-sem}.
In the $Soundness.lean$ file, the interested reader can find the formalization of the soundness theorem (whose statement we mention in Section~\ref{kripke-sound}), along with an auxiliary lemma used in its proof.
Then, $CompletenessListUtils.lean$ groups together some utilitary lemmas about $Finset$s of formulas, which are useful when proving some completeness-related theorems.
The Kripke completeness theorem, presented in Section~\ref{kripke-compl}, preceded by the definitions and results from Section~\ref{disj}, are formalized in the $Completeness.lean$ file.
Finally, the Heyting algebras notions and necessary results are formalized in the $HeytingAlgebraUtils.lean$ file, while the algebraic semantics, culminating with its associated completeness theorem and the equivalence between the validity notions
can be found in $HeytingAlgebraSemantics.lean$.

Fragments of Lean proofs will be included in the presentation only if we consider they contain worth-mentioning
technical aspects, or, in some cases, in order to sketch the key proof-steps.
The full source code is almost 3300 lines long and is available online in \cite{trufas-code}.

\section{Intuitionistic Propositional Logic}
In this section, we proceed to describe the main aspects of our formalization.
For full theoretical details of the results and proofs, the interested reader may refer to \cite{trufas-thesis}.
\subsection{Language}\label{lang}
We first formalize the countable set of propositional variables, as a wrapper over the $Nat$ type.
Structures are used to define non-recursive inductive data types, containing only one constructor.
And this is also the case here: we can identify any propositional variable with a natural number, so it is convenient to
define the $Var$ type as a structure with a single field, specifying the index of the variable:\\
\begin{lstlisting}
    structure Var where
      val : Nat
\end{lstlisting}
We work with a language containing falsity ($\bot$), conjunction ($\si$), disjunction ($\sau$) and implication ($\to$) as
primitive logical connectives. Thus, it is natural to define formulas by means of an inductive type,
in which the first non-recursive constructor uses the above defined structure type
and simply encapsulates it in a $Formula$ term, the second is meant to construct falsity,
while the following recursive constructors correspond each to one of the primitive connectives:\\
\begin{lstlisting}
    inductive Formula where
    | var : Var → Formula
    | bottom : Formula
\end{lstlisting}
\begin{lstlisting}
    | and : Formula → Formula → Formula
    | or : Formula → Formula → Formula
    | implication : Formula → Formula → Formula
\end{lstlisting} 
For readability reasons, we introduce the standard Unicode symbol for falsity
and define infix notations for the binary connectives,
which are much more convenient to use than the S-expressions in which Lean displays
the constructors by default. Additionally, we define the derived connectives for equivalence,
negation and truth, along with their standard notations:\\
\begin{lstlisting}
    notation "⊥" => bottom
    infixl:60 " ∧∧ " => and
    infixl:60 " ∨∨ " => or
    infixr:50 (priority := high) " ⇒ " => implication

    def equivalence (φ ψ : Formula) := (φ ⇒ ψ) ∧∧ (ψ ⇒ φ)
    infix:40 " ⇔ " => equivalence

    def negation (φ : Formula) : Formula := φ ⇒ ⊥
    prefix:70 " ~ " => negation

    def top : Formula := ~⊥
    notation " ⊤ " => top
\end{lstlisting}

\subsection{Proof system}\label{prf}
In this formalization, we adhere to the Hilbert-style proof system for IPL introduced
by Gödel in \cite{godel}. We define this using again an
inductive type, with constructors for each axiom and deduction rule:\\
\begin{lstlisting}
    inductive Proof (Γ : Set Formula) : Formula → Type where
    | premise {φ} : φ ∈ Γ → Proof Γ φ
    | contractionDisj {φ} : Proof Γ (φ ∨∨ φ ⇒ φ)
    | contractionConj {φ} : Proof Γ (φ ⇒ φ ∧∧ φ)
    | weakeningDisj {φ ψ} : Proof Γ (φ ⇒ φ ∨∨ ψ)
    | weakeningConj {φ ψ} : Proof Γ (φ ∧∧ ψ ⇒ φ)
    | permutationDisj {φ ψ} : Proof Γ (φ ∨∨ ψ ⇒ ψ ∨∨ φ)
    | permutationConj {φ ψ} : Proof Γ (φ ∧∧ ψ ⇒ ψ ∧∧ φ)
    | exfalso {φ} : Proof Γ (⊥ ⇒ φ)
    | modusPonens {φ ψ} : Proof Γ φ → Proof Γ (φ ⇒ ψ) → Proof Γ ψ
    | syllogism {φ ψ χ} : Proof Γ(φ ⇒ ψ) → Proof Γ(ψ ⇒ χ) → Proof Γ(φ ⇒ χ)
    | exportation {φ ψ χ} : Proof Γ (φ ∧∧ ψ ⇒ χ) → Proof Γ (φ ⇒ ψ ⇒ χ)
    | importation {φ ψ χ} : Proof Γ (φ ⇒ ψ ⇒ χ) → Proof Γ (φ ∧∧ ψ ⇒ χ) 
    | expansion {φ ψ χ} : Proof Γ (φ ⇒ ψ) → Proof Γ (χ ∨∨ φ ⇒ χ ∨∨ ψ)
\end{lstlisting}
The notion of $\Gamma$-theorem is defined as usual and we denote this by $\Gamma\vdash\vp$.
In Lean, we introduce this notation, as follows:\\
\begin{lstlisting}
    infix:25 " ⊢ " => Proof
\end{lstlisting}
The above definition of $Proof$ generates an elimination rule for this type,
which provides us with the formalized mechanisms of the recursion and induction principles
on proof terms.\\
Below we provide an example of how a pen-and-paper formal proof of a derived deduction rule can be transposed into
a mechanized Lean proof:\\
\begin{center}
\begin{tabular}{lll}
  (1) & $\Gamma\vdash\vp\si\psi\to\vp$  & (WEAKENING)\\[1mm]
  (2) & $\Gamma\vdash\vp\to\vp\sau\gamma$ & (WEAKENING)\\[1mm]
  (3) & $\Gamma\vdash\vp\si\psi\to\vp\sau\gamma$ & (SYLLOGISM): (1), (2)\\[2mm]
\end{tabular}
\end{center}
\begin{center}
\begin{tabular}{c}
\begin{lstlisting}
    def disjOfAndElimLeft : Γ ⊢ φ ∧∧ ψ ⇒ φ ∨∨ γ :=
      syllogism weakeningConj weakeningDisj
\end{lstlisting}
\end{tabular}
\end{center}
Note that, in the reverse-Hilbert formalized proof, we don't need to pass them explicitly,
when constructing the proof term, as the arguments of the constructors in the $Proof$ type
are implicit, so the Lean kernel will synthesize them from the context.

\subsection{Disjunctive theories, consistent and complete pairs}\label{disj}
These notions of disjunctive theories, consistent and complete pairs, and some results regarding them are essential in the Kripke completeness proof for IPL,
as we will see in Section \ref{kripke-compl}. Let us recall the definitions of these notions,
which can be consulted in \cite{kuznetsov}.
A set of formulas is said to be a disjunctive theory if it is deductively closed ($\Gamma\vdash\vp$ implies $\vp\in\Gamma$), consistent ($\Gamma\nvdash\bot$) and disjunctive ($\Gamma\vdash\vp\sau\psi$ implies $\Gamma\vdash\vp$ or $\Gamma\vdash\psi$).
Then, a pair of sets of formulas $(\Gamma, \Delta)$ is called consistent if there are no $G_1,\ldots,G_n\in\Gamma$ and $D_1,\ldots,D_m\in\Delta$, such that
$\vdash G_1\si\ldots\si G_n\to D_1\sau\ldots\sau D_m$.
Finally, we say that a consistent pair is complete, if it is a partition of the set of formulas.\\
\begin{lstlisting}
    def dedClosed {Γ : Set Formula} := ∀ (φ : Formula), Γ ⊢ φ → φ ∈ Γ

    def consistent {Γ : Set Formula} := Γ ⊢ ⊥ → False

    def disjunctive {Γ : Set Formula} :=
        ∀ (φ ψ : Formula), Γ ⊢ φ ∨∨ ψ → Sum (Γ ⊢ φ) (Γ ⊢ ψ)

    def disjunctiveTheory {Γ : Set Formula} :=
        @dedClosed Γ /\ @consistent Γ /\ Nonempty (@disjunctive Γ)

    def consistentPair {Γ Δ : Set Formula} :=
        ∀ (Φ Ω : Finset Formula), Φ.toSet ⊆ Γ → Ω.toSet ⊆ Δ →
        (∅ ⊢ Φ.toList.foldr Formula.and (~⊥) ⇒ Ω.toList.foldr Formula.or ⊥ →
        False)

    def completePair {Γ Δ : Set Formula} :=
        @consistentPair Γ Δ /\ ∀ (φ : Formula),(φ ∈ Γ /\ φ ∉ Δ) ∨ (φ ∈ Δ /\
        φ ∉ Γ)
\end{lstlisting}
Below we give the formalized statement of the lemma claiming that given a consistent pair,
any formula can be added to one of the sets in the pair, preserving the consistency:\\
\begin{lstlisting}
    lemma add_preserves_cons :
        @consistentPair Γ Δ → ∀ (φ : Formula), @consistentPair ({φ} ∪ Γ) Δ ∨
                                                 @consistentPair Γ ({φ} ∪ Δ)
\end{lstlisting}
The proof of the above lemma follows by reductio ad absurdum and it requires a syntactical derivation,
but it doesn't give rise to any technical difficulties, so we do not present it here.\\
Then, to prove the essential $consistent\_incl\_complete$ lemma, stating that any consistent pair can be component-wise included in a complete one, we define an
indexed family of formula-set pairs, thus:\\
\begin{lstlisting}
    def family (nf : Nat → Formula) (n : Nat) : Set Formula × Set Formula :=
        match n with
\end{lstlisting}
\begin{lstlisting}
        | .zero => @add_formula_to_pair Γ Δ (nf 0)
        | .succ n => @add_formula_to_pair (family nf n).fst (family nf n).snd
                     (nf (n + 1))
\end{lstlisting}
To have access to an enumeration of formulas, we pass as the first argument
a function which assigns, to any natural number, a formula.
Then, we inductively build the family, by adding the formulas to one of the sets in the pair,
whilst preserving the consistency. Without loss of generality, we define
the function to add the formula to the first set in the pair, if possible:\\
\begin{lstlisting}
    def add_formula_to_pair (φ : Formula) : Set Formula × Set Formula :=
        if @consistentPair ({φ} ∪ Γ) Δ then (({φ} ∪ Γ), Δ)
        else (Γ, {φ} ∪ Δ)
\end{lstlisting}
By the $add\_preserves\_cons$ lemma previously presented, it follows easily that applying the above defined
$add\_formula\_to\_pair$ function repeatedly, starting from a consistent pair, we preserve the consistency
of the obtained pairs.\\

\noindent
The enumeration of formulas is not required to be bijective, a surjection from
$Nat$ to $Formula$ is sufficient in this case, as we don't have any restriction for adding the formulas only once.
Classically, the existence of an injective function from a type $\alpha$ to a type $\beta$
gives evidence that there is a surjection from $\beta$ to $\alpha$.
Hence, we define an injective function from $Formula$ to $Nat$.\\
To construct the injection, we use Cantor's pairing function, which we multiply by two,
for ease of formalization. For a theretical presentation of Cantor's encoding, refer to Section 1.3.9 in \cite{troelstra}.\\
\begin{lstlisting}
    def pairing (x y : ℕ) := (x + y) * (x + y + 1) + 2 * x
\end{lstlisting}
Then, we associate a numerical identifier to any connective symbol and
encode formulas into natural numbers by recursively applying the pairing function
on the structure of the formula, as follows:\\
\begin{lstlisting}
    def encode_form : Formula → ℕ
    | var v => pairing 0 (v.val + 1)
    | bottom => 0
    | φ ∧∧ ψ => pairing (pairing (encode_form φ) 1) (encode_form ψ)
    | φ ∨∨ ψ => pairing (pairing (encode_form φ) 2) (encode_form ψ)
    | φ ⇒ ψ => pairing (pairing (encode_form φ) 3) (encode_form ψ)
\end{lstlisting}
After proving the injectivity of our encoding function, we are able to define
an instance of $Countable$ for our $Formula$ type. The Mathlib definition of the $Countable$ type-class is
as follows:\\
\begin{lstlisting}
    class Countable (α : Sort u) : Prop where
        exists_injective_nat' : ∃ f : α → ℕ, Injective f
\end{lstlisting}
So we immediately define the $Countable$ instance for the $Formula$ type, based on the proof of the encoding's injectivity:\\
\begin{lstlisting}
    instance : Countable Formula := inject_Form.countable
\end{lstlisting}
Now, having the surjective enumeration at hand, we can get a step closer to the
final construction of the complete pair which includes the initial consistent pair component-wise. We prove
that any formula $\vp$ is contained in one of the sets of the pair with index $fn(\vp)$, where
by $fn$ we denote the injective encoding of formulas into natural numbers:\\
\begin{lstlisting}
    lemma vp_in_ΓiΔi (φ : Formula) (fn : Formula → Nat) (fn_inj : fn.Injective)
      (nf : Nat → Formula) (nf_inv : nf = fn.invFun) :
       φ ∈ (@family Γ Δ nf (fn φ)).fst \/ φ ∈ (@family Γ Δ nf (fn φ)).snd
\end{lstlisting}
In Mathlib, the inverse of a function is noncomputably defined as follows:\\
\begin{lstlisting}
    noncomputable def invFun {α : Sort u} {β} [Nonempty α] (f : α → β) :
        β → α :=
        fun y ↦ if h : (∃ x, f x = y) then h.choose else Classical.arbitrary α
\end{lstlisting}
So this is why we can count on this inverse for any function, regardless of its bijectivity.
Notice also that the injectivity of $fn$ gives evidence of $invFun$ being the so-called $left-inverse$.\\
It is also crucial to prove that the family we defined is increasing:\\
\begin{lstlisting}
    lemma increasing_family {nf : Nat → Formula} (i j : Nat) : i <= j →
        (@family Γ Δ nf i).fst ⊆ (@family Γ Δ nf j).fst /\
        (@family Γ Δ nf i).snd ⊆ (@family Γ Δ nf j).snd
\end{lstlisting}
Next, we define the component-wise union of the indexed pair-family:\\
\begin{lstlisting}
    def consistent_family_union (_ : @consistentPair Γ Δ) (nf : Nat → Formula) :
        Set Formula × Set Formula :=
        ({φ | ∃ i : Nat, φ ∈ (@family Γ Δ nf i).fst},
         {φ | ∃ i : Nat, φ ∈ (@family Γ Δ nf i).snd})
\end{lstlisting}
This is finally the witness we make use of when proving the existence of a complete pair,
component-wise including our initial consistent one. Of course, before using the family union
this way, we have to give evidence that it is indeed a partition of the set of formulas.
The increasing property is crucial in achieving this last-mentioned goal.\\
Finally, we present the formalized statement of the $consistent\_incl\_complete$ lemma:\\
\begin{lstlisting}
    lemma consistent_incl_complete :
      @consistentPair Γ Δ → (∃ (Γ' Δ' : Set Formula), Γ ⊆ Γ' ∧ Δ ⊆ Δ' ∧
      @completePair Γ' Δ')
\end{lstlisting}
This will be useful when proving the completeness of IPL with respect to the Kripke semantics,
which will be subsequently presented.

\subsection{Kripke semantics}\label{kripke-sem}
In the sequel, we define the Kripke semantics.
The first definition we need is, of course, that of a Kripke model. We first state this informally, then
provide its corresponding formalization. An intuitionistic propositional Kripke model is a tuple $(W, R, V)$, where $W$ is a non-empty set,
$R$ is a reflexive and transitive binary relation on W and $V:Var\times W\ra\{0,1\}$ is a function assigning truth values to variables.
$V$ is assumed to be monotone with respect to R, thus
$V(p, w)=1$ and $Rww'$ implies $V(p, w')=1$.\\
\begin{lstlisting}
    structure KripkeModel (W : Type) where
        R : W → W → Prop
        V : Var → W → Prop
        refl (w : W) : R w w
        trans (w1 w2 w3 : W) : R w1 w2 → R w2 w3 → R w1 w3
        monotonicity (v : Var) (w1 w2 : W) : R w1 w2 → V v w1 → V v w2
\end{lstlisting}
We formalize the Kripke model as a parameterized structure, where the
parameter W represents the space of worlds. Thus, the worlds of a model
are in Lean terms of type W. The first field of the structure models the accessibility
binary relation R over terms of type W and V is the valuation function, which takes
two arguments - a variable and an inhabitant of type W. Then, the last three
fields are meant to formalize the properties of the relation R
(reflexivity and transitivity) and the monotonicity of the valuation.\\

\noindent
The extended valuation function (on formulas) is defined as follows:\\
\begin{lstlisting}
    def val {W : Type} (M : KripkeModel W) (w : W) : Formula → Prop
    | Formula.var p => M.V p w
    | ⊥ => False
    | φ ∧∧ ψ => val M w φ /\ val M w ψ
    | φ ∨∨ ψ => val M w φ \/ val M w ψ
    | φ ⇒ ψ => ∀ (w' : W), M.R w w' /\ val M w' φ → val M w' ψ
\end{lstlisting}
We say that a formula $\vp$ is true at a world $w$ of a model $M$, if $V(\vp,w)=1$ and we denote this by $M,w\vDash\vp$.
Then, $\vp$ is said to be valid in a model $M:=(W,R,V)$, if $M,w\vDash\vp$, for all $w\in W$.
And finally, $\vp$ is valid, if it is valid in all the Kripke models. We denote this by $\vDash\vp$.\\
Below, we present the formalization of these notions:\\
\begin{lstlisting}
    def true_in_world {W : Type} (M : KripkeModel W) (w : W) (φ : Formula): Prop :=
        val M w φ

    def valid_in_model {W : Type} (M : KripkeModel W) (φ : Formula) : Prop :=
        ∀ (w : W), val M w Φ

    def valid (φ : Formula) : Prop :=
        ∀ (W : Type) (M : KripkeModel W), valid_in_model M Φ
\end{lstlisting}
We say that $M, w$ forces $\Gamma$ (and denote it by $M,w\vDash\Gamma$), if $M,w\vDash\vp$, for all $\vp\in\Gamma$.\\
\begin{lstlisting}
    def model_sat_set {W : Type}(M : KripkeModel W)(Γ : Set Formula)(w : W):Prop:=
        ∀ (φ : Formula), φ ∈ Γ → val M w φ
\end{lstlisting}
Another essential notion is that of local semantic consequence.
We say that a formula $\vp$ is a local semantic consequence of a set $\Gamma$,
if for all models $M$, and all worlds $w$ in M, we have that $M, w\forces\Gamma$ implies $M, w\forces\vp$.
We denote this by $\Gamma\vDash\vp$.\\
\begin{lstlisting}
    def sem_conseq (Γ : Set Formula) (φ : Formula) : Prop :=
        ∀ (W : Type) (M : KripkeModel W) (w : W),
        model_sat_set M Γ w → val M w φ
    infix:50 " ⊨ " => sem_conseq
\end{lstlisting}
Then, a set $\Delta$ is forced by $\Gamma$, if $\Gamma\vDash\vp$, for all $\vp$ in $\Delta$.\\
\begin{lstlisting}
    def set_forces_set (Γ Δ : Set Formula) : Prop :=
        ∀ (φ : Formula), φ ∈ Δ → Γ ⊨ φ
\end{lstlisting}

\subsection{Kripke completeness theorem}\label{kripke-sound-compl}
\subsubsection{Soundness}\label{kripke-sound}
The soundness theorem claims that any $\Gamma$-theorem is a local semantic consequence of $\Gamma$
($\Gamma\vdash\vp$ implies $\Gamma\vDash\vp$), for any set of formulas $\Gamma$ and any formula $\vp$.\\
In Lean, this statement transposes to:\\
\begin{lstlisting}
    theorem soundness (Γ : Set Formula) (φ : Formula) : Γ ⊢ φ → Γ ⊨ φ
\end{lstlisting}
The proof is straightforward, so we briefly sketch it here.
For full detail, the interested reader shall consult the formalization.\\
We proceed by induction on $Proof$. For all the axiom cases, we apply an auxiliary lemma asserting that
any axiom is valid:\\
\begin{lstlisting}
    lemma axioms_valid (φ : Formula) (ax : Axiom φ) : valid φ
\end{lstlisting}
Worth-mentioning is also the use of the monotonicity property of the valuation function,
in the $exportation$ case. We prove this result in $Semantics.lean$ and mention here only
its formalized claim:\\
\begin{lstlisting}
    lemma monotonicity_val (W : Type) (M : KripkeModel W) (w1 w2 : W) (φ : Formula):
        M.R w1 w2 → val M w1 φ → val M w2 φ
\end{lstlisting}
\subsubsection{Completeness}\label{kripke-compl}
\begin{theorem}\label{completeness}(completeness theorem)
For any set of formulas $\Gamma$ and any formula $\vp$:
\begin{center}
$\Gamma\vdash\vp$ iff $\Gamma\forces\vp$.\\
\end{center}
\end{theorem}
\noindent
The left implication is the soundness theorem, which was already proved in Section~\ref{kripke-sound}.
\noindent
For the reverse implication in the completeness theorem, we appeal to nonconstructive reasoning,
proceeding by contraposition. More precisely, we assume by reductio ad absurdum that $\Gamma\nvdash\vp$
and then construct a Kripke model (the so-called canonical model), which satisfies $\Gamma$, but does not
satisfy $\vp$. Hence, we get that $\vp$ is not a local semantic consequence of $\Gamma$, which contradicts
our assumption. Our approach follows the Henkin-style completeness proof presented in \cite{kuznetsov}.\\

\noindent
We first describe the construction of the canonical model. The domain is set to the type of the disjunctive theories.
This $setDisjTh$ type is defined as a subtype of the $Set\: Formula$ type, as follows:\\
\begin{lstlisting}
    abbrev setDisjTh := {Γ // @disjunctiveTheory Γ}
\end{lstlisting}
For the $refl$, $trans$, and $monotonicity$ fields of the structure, we have to pass proofs of the
set inclusion relation satisfying these properties. These proofs are easily
completed, using the corresponding Mathlib theorems. Putting this all together, we have:\\
\begin{lstlisting}
    def canonicalModel : KripkeModel (setDisjTh) :=
    {
        R := fun (Γ Δ) => Γ.1 ⊆ Δ.1,
        V := fun (v Γ) => Formula.var v ∈ Γ.1,
        refl := fun (Γ) => Set.Subset.rfl
        trans := fun (Γ Δ Φ) => Set.Subset.trans
        monotonicity := fun (v Γ Δ) => by intros; apply Set.mem_of_mem_of_subset
                                        assumption'
    }
\end{lstlisting}
Apart from lemma $consistent\_incl\_complete$ we have already presented in Section \ref{disj}, the Kripke completeness
proof requires also the so-called main semantic lemma. This lemma states that the property of the valuation
in the definition of the canonical model, holds also for the extended valuation function on formulas. Thus, it claims that
$M_0,\Gamma\vDash\vp$ if and only if $\vp\in\Gamma$, for any disjunctive theory $\Gamma$ and formula $\vp$:\\
\begin{lstlisting}
    lemma main_sem_lemma (Γ : setDisjTh) (φ : Formula) :
        val canonicalModel Γ φ ↔ φ ∈ Γ.1
\end{lstlisting}
It is worth mentioning that the two implications in this lemma cannot be formalized as independent lemmas,
because of the $implication$ case, where the proof of the left implication depends on the
right implication in the induction hypothesis, and vice versa.\\
Now we have all the necessary ingredients for the completenss contraposition proof informally presented at the beginning of this section.
The formalized completeness statement
is the following:\\
\begin{lstlisting}
    theorem completeness {φ : Formula} {Γ : Set Formula} :
        Γ ⊨ φ ↔ Nonempty(Γ ⊢ φ)
\end{lstlisting}

\subsection{Algebraic semantics and completeness theorem}\label{alg}
Our approach in the current section is based on the exposition in the textbook \cite{rasiowa} and the lecture notes \cite{georgescualgebras,georgesculogic}.
After establishing the Heyting algebras necessary premises, we
move on to defining the algebraic models of IPL and the Lindenbaum-Tarski algebra. Finally, we provide a
second completeness proof, with respect to the algebraic semantics and prove the equivalence between the
Kripke and algebraic validity.
\subsubsection{Heyting algebras}
First of all, we shall recall the definition of a Heyting algebra. A Heyting algebra (or pseudo-boolean algebra) is a structure $(H,\vee,\wedge,\to)$ such that $H$ is a bounded lattice and the following residuation property holds:
$a\le b\to c$ if and only if $a\si b\le c$. Conventionally, we denote a Heyting algebra by $H$.\\
We start by formalizing the general definitions on Heyting algebras.
Mathlib contains a definition of the $HeytingAlgebra$ type class, which encompasses the conditions a type has to satisfy,
in order to have the structure of a Heyting algebra. However, we have to formalize and prove
the necessary definitions and results about filters.\\
We consider a type $\alpha$ for which there is an instance of the Mathlib $HeytingAlgebra$ class:\\
\begin{lstlisting}
    variable {α : Type u} [HeytingAlgebra α]
\end{lstlisting}
Then, we formalize the following main definitions, using the above $\alpha$ type-variable,
to represent the domain of the Heyting algebra.\\
A filter is a nonempty set $F$, satisfying two conditions: (i)
for any $x, y\in F$, $x\wedge y\in F$, (ii) for any $x\in F$ and $y\ge x$,
we have that $y\in F$.\\
\begin{lstlisting}
    def filter (F : Set α) := (Set.Nonempty F) ∧ (∀ (x y : α), x ∈ F → y ∈ F →
                               x ⊓ y ∈ F) ∧ (∀ (x y : α), x ∈ F → x ≤ y → y∈F)
\end{lstlisting}
The filter generated by a set $X$ is the intersection of all the filters which include $X$.\\
\begin{lstlisting}
    abbrev X_filters (X : Set α) := {F // filter F ∧ X ⊆ F}
    def X_gen_filter (X : Set α) := {x | ∀ (F : X_filters X), x ∈ F.1}
\end{lstlisting}
A filter is called proper, if it doesn't contains the first element of the lattice.\\
\begin{lstlisting}
    def proper_filter (F : Set α) := filter F ∧ ⊥ ∉ F
\end{lstlisting}
Additionally, a proper filter $F$ is said to be prime, if for all $x,y\in H$, if $x\sau y\in F$, then $x\in F$ or $y\in F$.\\
\begin{lstlisting}
    def prime_filter {α : Type} [HeytingAlgebra α] (F : Set α) :=
        proper_filter F ∧ (∀ (x y : α), x ⊔ y ∈ F → x ∈ F ∨ y ∈ F)
\end{lstlisting}
Next, we present the central Heyting algebras result, which will be used in a subsequent section,
when transiting from an algebraic model to the corresponding Kripke one.
It asserts that, given a filter $F$ and an element $x$ which is not in F, there exists a prime filter $P$ including the
initial filter, such that $x$ is neither an element of $P$:\\
\begin{lstlisting}
    lemma super_prime_filter (x : α) (F : Set α) (Hfilter : @filter α _ F)
       (Hnotin : x ∉ F) :
        ∃ (P : Set α), @prime_filter α _ P /\ F ⊆ P /\ x ∉ P
\end{lstlisting}
In the following, we informally sketch the proof of the above lemma and present key-fragments of its formalization.
First of all, we show that the set of all the prime filters not containing $x$
has an upper bound:\\
\begin{lstlisting}
    have Hzorn : ∃ F' ∈ X_filters_not_cont_x x, F ⊆ F' ∧
                 ∀ (F'' : Set α), F'' ∈ X_filters_not_cont_x x → F' ⊆ F'' →
                    F'' = F'    
\end{lstlisting}
This is achieved by applying Zorn's lemma, which is formalized in Mathlib as follows :\\
\begin{lstlisting}
    theorem zorn_subset_nonempty (S : Set (Set α))
        (H : ∀ (c) (_ : c ⊆ S), IsChain (· ⊆ ·) c → c.Nonempty →
             ∃ ub ∈ S, ∀ s ∈ c, s ⊆ ub) (x)
        (hx : x ∈ S) : ∃ m ∈ S, x ⊆ m ∧ ∀ a ∈ S, m ⊆ a → a = m
\end{lstlisting}
where $isChain$ is a $Prop$ deciding whether a given set is totally ordered.
The upper bound we are looking for is the union of all the chain's elements.
In the rest of the proof, our goal is to prove that this upper bound is a prime filter,
and we proceed by contraposition, in doing so. We consider two elements $y,z$
such that $y\notin P$ and $z\notin P$. Then, the first step is showing that
$P\subset[P\cup\{y\})$ and its analogous $P\subset[P\cup\{z\})$.
Using these auxiliary hypotheses and the maximality of $P$, we prove that $x\in[P\cup\{y\})$ and $x\in[P\cup\{z\})$.
Now, having also this hypothesis at hand, the proof concludes by applying a few well-known Heyting algebras properties,
as already shown in the theoretical proof.\\

\noindent
The following lemma provides a useful characterization of the filter generated by a set $X$:\\
\begin{lstlisting}
    lemma gen_filter_prop (X : Set α) :
        X_gen_filter X = {a | ∃ (l : List α), l.toFinset.toSet ⊆ X∧inf_list l≤a}
\end{lstlisting}
We use this form of the generated filter to obtain an auxiliary result which is
necessary for the proof of the above $super\_prime\_filter$ lemma:\\
\begin{lstlisting}
    lemma mem_gen_ins_filter (F : Set α) (Hfilter : filter F) :
        y ∈ X_gen_filter (F ∪ {x}) → ∃ (z : α), z ∈ F /\ x ⊓ z ≤ y
\end{lstlisting}
Applying this last lemma, the residuation property and a few basic properties of Heyting algebras and filters, we obtain another important result, which will be used when constructing
the valuation function of the Kripke model associated to an algebraic one:\\
\begin{lstlisting}
    lemma himp_not_mem (F : Set α) (Hfilter : filter F) (Himp_not_mem : x ⇒ y∉F) :
        y ∉ X_gen_filter (F ∪ {x})
\end{lstlisting}
The $super\_prime\_filter$ lemma has also a couple of corollaries. The first one states that given an element $x$
different from the last element of the algebra, there exists a prime filter $P$
such that $x\notin P$:\\
\begin{lstlisting}
    lemma super_prime_filter_cor1 (x : α) (Hnottop : x ≠ ⊤) :
        ∃ (P : Set α), @prime_filter α _ P /\ x ∉ P
\end{lstlisting}
To prove this, we trivially show first that $\{\top\}$ is a filter and then, using the $super\_prime\_filter$ lemma,
we obtain the necessary witness.\\
The second corollary follows immediately from the first one. It claims that
intersecting all the prime filters, we obtain the set $\{\top\}$:\\
\begin{lstlisting}
    lemma super_prime_filter_cor2 : Set.sInter (@prime_filters α _) = {⊤} :=
\end{lstlisting}
This is proved by double inclusion and will be of great importance in an upcoming section, when establishing the connection
between the two semantical paradigms.

\subsubsection{Algebraic models}
An algebraic interpretation in $H$ is a function $\algint:Form\to H$ satisfying the 
following conditions: $\algint(\bot)=0$ and, for all $\vp,\psi\in Form$,
$\algint(\vp\si\psi)=\algint(\vp)\si \algint(\psi)$,
$\algint(\vp\sau\psi)=\algint(\vp)\sau \algint(\psi)$ and
$\algint(\vp\to\psi)=\algint(\vp)\ra \algint(\psi)$.\\
We formalize the notion of algebraic interpretation as follows:\\
\begin{lstlisting}
    def AlgInterpretation (I : Var → α) : Formula → α
    | Formula.var p => I p
    | Formula.bottom => ⊥
    | φ ∧∧ ψ => AlgInterpretation I φ ⊓ AlgInterpretation I ψ
    | φ ∨∨ ψ => AlgInterpretation I φ ⊔ AlgInterpretation I ψ
    | φ ⇒ ψ => AlgInterpretation I φ ⇒ AlgInterpretation I ψ
\end{lstlisting}
An algebraic model is a tuple $(H,\algint)$.\\
We've chosen not to explicitly define the notion of algebraic model in Lean, since
it would have implied to adjoin the above defined interpretation function to the type.
We considered this redundant, since an algebraic model is uniquely determined by the
variable-interpretation function.\\
A formula $\vp$ is true in an algebraic model $(H,\algint)$, if $\algint(\vp)=1$. We denote this by $(H,\algint)\forcesalg\vp$.
We say that $\vp$ is algebraically valid in $H$, if $(H,\algint)\forcesalg\vp$, for any algebraic model $(H,\algint)$.
Finally, $\vp$ is called algebraically valid, if $\vp$ is algebraically valid in any Heyting algebra $H$.
This is denoted by $\forcesalg\vp$.\\
\begin{lstlisting}
    def true_in_alg_model (I : Var → α) (φ : Formula) : Prop :=
        AlgInterpretation I φ = Top.Top

    def valid_in_alg (φ : Formula) : Prop :=
        ∀ (I : Var → α), true_in_alg_model I φ

    def alg_valid (φ : Formula) : Prop :=
        ∀ (α : Type) [HeytingAlgebra α], @valid_in_alg α _ φ
\end{lstlisting}
A set of formulas $\Gamma$ is true in an algebraic model, if  
$(H,\algint)\forcesalg\vp$ for any $\vp\in\Gamma$. We denote this by $(H,\algint)\forcesalg\Gamma$. We say that $\Gamma$
is algebraically valid in $H$, if $(H,\algint)\forcesalg\Gamma$, for any algebraic model $(H,\algint)$.
A set $\Gamma$ is algebraically valid, if it is algebraically valid in any Heyting algebra $H$.
This is denoted by $\forcesalg\Gamma$.\\
\begin{lstlisting}
    def set_true_in_alg_model (I : Var → α) (Γ : Set Formula) : Prop :=
        ∀ (φ : Formula), φ ∈ Γ → AlgInterpretation I φ = Top.top
\end{lstlisting}
\vspace{\baselineskip}
\begin{lstlisting}
    def set_valid_in_alg (Γ : Set Formula) : Prop :=
        ∀ (I : Var → α), set_true_in_alg_model I Γ
\end{lstlisting}
\vspace{\baselineskip}
\begin{lstlisting}
    def set_alg_valid (Γ : Set Formula) : Prop :=
        ∀ (α : Type) [HeytingAlgebra α], @set_valid_in_alg α _ Γ
\end{lstlisting}
We say that $\vp$ is an \algsemcons of $\Gamma$, if for any algebraic model $(H,\algint)$,
$(H,\algint) \forcesalg \Gamma$ implies $(H,\algint) \forcesalg \vp$.
We denote this by $\Gamma\forcesalg\vp$.\\
\begin{lstlisting}
    def alg_sem_conseq (Γ : Set Formula) (φ : Formula) : Prop :=
        ∀ (α : Type)[HeytingAlgebra α](I : Var → α), set_true_in_alg_model I Γ →
        true_in_alg_model I φ
    infix:50 " ⊨ₐ " => alg_sem_conseq
\end{lstlisting}

\subsubsection{Lindenbaum-Tarksi algebra}\label{lt}
We define the following equivalence relation on formulas, with respect to a set $\Gamma$:
\begin{center}
$\vp\sim_\Gamma\psi$ iff $\Gamma\vdash\vp\dnd\psi$
\end{center}
Let $Form/\sim_\Gamma$ be the quotient set.
We denote the equivalence class of a formula $\vp$ by $\equivcGamma{\vp}$.
The order relation on $Form/\sim_\Gamma$
is defined as follows:
$\equivcGamma{\vp}\le_\Gamma\equivcGamma{\psi}$ iff $\Gamma\vdash\vp\to\psi$.\\
Then, the quotient set $Form/\sim_\Gamma$ is a Heyting algebra (called the Lindenbaum-Tarksi algebra), where:
$\equivcGamma{\vp}\sau\equivcGamma{\psi}=\equivcGamma{\vp\sau\psi}$,
$\equivcGamma{\vp}\si\equivcGamma{\psi}=\equivcGamma{\vp\si\psi}$, 
$\equivcGamma{\vp}\to\equivcGamma{\psi}=\equivcGamma{\vp\to\psi}$,
$\equivcGamma{\bot}$ is the first element and
$\equivcGamma{\neg\bot}$ is the last element.\\
First of all, we formalize the equivalence relation on formulas
with respect to $\Gamma$, along with its standard infix notation:\\
\begin{lstlisting}
    def equiv (φ ψ : Formula) := Nonempty (Γ ⊢ φ ⇔ ψ)
    infix:50 "~" => equiv
\end{lstlisting}
Next, we define a setoid instance for our $Formula$ type, by providing a proof
of the above defined relation being indeed an equivalence relation and then
we can move to defining the $\le,\si,\sau,\to$ operations on quotients of this setoid.
To define quotient conjunction, disjunction and implication, we make use
of the built-in $lift_2$ function, which lifts the corresponding binary functions on formulas,
to a quotient on both arguments. We give below only the formalization of
quotient conjunction. The other quotient operations are defined in a similar manner.\\
\begin{lstlisting}
    def Formula.and_quot (φ ψ : Formula) := Quotient.mk setoid_formula (φ ∧∧ ψ)

    def and_quot (φ ψ : Quotient setoid_formula) : Quotient setoid_formula :=
        Quotient.lift₂ Formula.and_quot and_quot_preserves_equiv φ ψ
\end{lstlisting}
Notice the fact that we have to pass as the second argument of $lift_2$ a
proof of our binary operation preserving equivalence. The statement of the corresponding
lemma is as follows:\\
\begin{lstlisting}
    lemma and_quot_preserves_equiv (φ ψ φ' ψ' : Formula) : φ ~ φ' → ψ ~ ψ' →
        (Formula.and_quot φ ψ = Formula.and_quot φ' ψ')
\end{lstlisting}
Having these operations defined, we can prove that the quotient type associated
to the $\sim$ equivalence relation is a Heyting algebra. We do so by defining
a Heyting algebra instance for this type:\\
\begin{lstlisting}
    instance lt_heyting : HeytingAlgebra (Quotient (@setoid_formula Γ))
\end{lstlisting}
We don't provide the full definition of this instance here, but all the proofs
we need to complete its fields are rather trivial.\\
We define the mapping which associates to a formula its corresponding quotient:\\
\begin{lstlisting}
    def h_quot_var (v : Var) : Quotient (@setoid_formula Γ) :=
        Quotient.mk setoid_formula (Formula.var v)
\end{lstlisting}
\vspace{\baselineskip}
\begin{lstlisting}
    def h_quot (φ : Formula) : Quotient (@setoid_formula Γ) :=
        Quotient.mk setoid_formula φ
\end{lstlisting}
The $h\_quot\_var$ function will be passed as an argument to $AlgInterpretation$,
when proving that $h\_quot$ satisfies the conditions of an algebraic interpretation.
The statement of this lemma is as follows:\\
\begin{lstlisting}
    lemma h_quot_interpretation : ∀ (φ : Formula),  h_quot φ = (@AlgInterpretation (Quotient (@setoid_formula Γ)) _ h_quot_var φ)
\end{lstlisting}
Then, we are able to prove the two results about the Lindenbaum-Tarski algebra, which will be crucial
in the proof of the algebraic completeness theorem. The first one asserts that a set $\Gamma$
is true at the algebraic model generated by itself, whilst the second claims that a formula $\vp$
is true at the algebraic model induced by $\Gamma$, if and only if $\vp$ is a $\Gamma$-theorem.
We mention only their statements below, as the proofs do not contain any
technical difficulties:\\
\begin{lstlisting}
    lemma set_true_in_lt :
        @set_true_in_alg_model (Quotient (@setoid_formula Γ)) _ h_quot_var Γ
\end{lstlisting}
\vspace{\baselineskip}
\begin{lstlisting}
    lemma true_in_lt (φ : Formula) :
        @true_in_alg_model (Quotient (@setoid_formula Γ)) _ h_quot_var φ ↔
        Nonempty (Γ ⊢ φ)
\end{lstlisting}

\subsubsection{Algebraic completeness theorem}
\bthm\label{alg-completeness}[algebraic completeness]
For any set of formulas $\Gamma$ and any formula $\vp$, 
\bce
$\Gamma\vdash \vp$ iff $\Gamma \forcesalg \vp$.
\ece
\ethm
\noindent
The soundness implication follows immediately, by a straightforward induction. We mention only its formalized statement here:\\
\begin{lstlisting}
    theorem soundness_alg (φ : Formula) : Nonempty (Γ ⊢ φ) → alg_sem_conseq Γ φ 
\end{lstlisting}
Moving now to the reverse implication, the proof is based on the two results
mentioned at the end of Section \ref{lt}. Below, we present the full formalization
of the algebraic completeness theorem:\\
\begin{lstlisting}
    theorem completeness_alg (φ : Formula) :
      alg_sem_conseq Γ φ ↔ Nonempty (Γ ⊢ φ) :=
        by
          apply Iff.intro
          · intro Halg
            rw [<-true_in_lt]
            exact Halg (Quotient (@setoid_formula Γ)) h_quot_var set_true_in_lt
          · exact soundness_alg φ
\end{lstlisting}

\subsubsection{Kripke models and algebraic models}
The central result in this last section is the equivalence between the two validity notions:
\begin{center}
$\forces\vp$ iff $\forcesalg\vp$
\end{center}
We follow the approach in \cite{fitting} and hence give a pure semantical proof of the above mentioned result, wihtout using the completeness theorems of the two semantics.
We start by establishing a connection from Kripke models to algebraic models.
In doing so, we have to define first the notions of closed set, and the Heyting algebra structure
which can be built on top of the set of all the closed sets.\\
Thus, the following $Prop$ decides whether a domain set of a Kripke model is closed:\\
\begin{lstlisting}
    def closed {W : Type} (M : KripkeModel W) (A : Set W) : Prop :=
        ∀ (w w' : W), w ∈ A → M.R w w' → w' ∈ A
\end{lstlisting}
We formalize the set of all closed subsets as a subtype of the $Set\,W$ type, as follows:\\
\begin{lstlisting}
    def all_closed {W : Type} (M : KripkeModel W) := {A // @closed W M A}
\end{lstlisting}
For the implication operation on closed subsets, we first define the set of all
closed sets contained in $W\setminus A\cup B$, where by $A,B$ we denote the two
implication operands. Then, the union of the elements in this set is the greatest
closed set satisfying our condition:\\
\begin{lstlisting}
    def all_closed_subset {W : Type} (M : KripkeModel W) (A B : all_closed M) :=
        {X | @closed W M X /\ X ⊆ ((@Set.univ W) \ A.1) ∪ B.1}
\end{lstlisting}
\vspace{\baselineskip}
\begin{lstlisting}
    def himp_closed {W : Type} {M : KripkeModel W} (A B : all_closed M) :=
        Set.sUnion (@all_closed_subset W M A B)
\end{lstlisting}
We define the corresponding Heyting algebra instance, as follows:\\
\begin{lstlisting}
    instance {W : Type} (M : KripkeModel W) : HeytingAlgebra (all_closed M) :=
      { sup := λ X Y => {val := X.1 ∪ Y.1, property := union_preserves_closed X Y}
        le := λ X Y => X.1 ⊆ Y.1
        le_refl := λ _ => Set.Subset.rfl
        le_trans := λ _ _ _ => Set.Subset.trans
        le_antisymm := λ _ _ => by rw [Subtype.ext_iff]; apply Set.Subset.antisymm
        le_sup_left := λ X Y => Set.subset_union_left X.1 Y.1
        le_sup_right := λ X Y => Set.subset_union_right X.1 Y.1
        sup_le := λ _ _ _ => Set.union_subset
        inf := λ X Y => {val := X.1 ∩ Y.1, property := inter_preserves_closed X Y}
        inf_le_left := λ X Y => Set.inter_subset_left X.1 Y.1
        inf_le_right := λ X Y => Set.inter_subset_right X.1 Y.1
        le_inf := λ _ _ _ => Set.subset_inter
        top := {val := @Set.univ W, property := univ_closed}
        le_top := λ X => Set.subset_univ X.1
        himp := λ X Y => {val := himp_closed X Y, property := himp_is_closed X Y}
        le_himp_iff := λ X Y Z => himp_closed_prop Y Z X
        bot := {val := ∅, property := empty_closed}
        bot_le := λ X => Set.empty_subset X.1
        compl := λ X => {val := himp_closed X {val := ∅, property := empty_closed},
                         property := himp_is_closed X {val := ∅,
                                                       property := empty_closed}}
        himp_bot := by simp }
\end{lstlisting}
The next step is proving that the following function is an algebraic interpretation:\\
\begin{lstlisting}
    def h {W : Type} {M : KripkeModel W} (φ : Formula) : all_closed M :=
\end{lstlisting}
\begin{lstlisting}
        {val := {w | val M w φ}, property := by intro w w' Hwin Hr
                                                 apply monotonicity_val
                                                 assumption'}
\end{lstlisting}
Except for the implication case, the proof is trivial.
We present here the main steps of this last interesting case.
The proof is by double inclusion, but before succeeding in doing so,
we need to prove an additional statement, which holds only for closed subsets:\\
\begin{lstlisting}
    have Haux : ∀ (A : all_closed M),
        A.1 ⊆ (@h W M (ψ ⇒ χ)).1 ↔ A.1 ∩ (@h W M ψ).1 ⊆ (@h W M χ).1
\end{lstlisting}
By this point, we can formalize the first central result of the section,
which provides a method of constructing an algebraic model corresponding to a given Kripke model:\\
\begin{lstlisting}
    lemma kripke_alg {W : Type} {M : KripkeModel W} (φ : Formula) :
      valid_in_model M φ ↔ @true_in_alg_model (all_closed M) _ h_var φ
\end{lstlisting}
In the sequel, we aim to formalize also the reverse direction, namely the switch from an algebraic model to a corresponding Kripke one.
We first define the Kripke frame based on the set of all prime filters.
The accessibility relation is given by inclusion and a variable is said to be true at a world of a prime filter $F$,
if it is an element of $F$:\\
\begin{lstlisting}
    def prime_filters_frame (I : Var → α) :
        KripkeModel (@prime_filters α _) :=
        {
            R := λ (F1 F2) => F1.1 ⊆ F2.1,
            V := λ (v F) => I v ∈ F.1,
            refl := λ (F) => Set.Subset.rfl,
            trans := λ (F1 F2 Φ) => Set.Subset.trans,
            monotonicity := λ (v F1 F2) => by intros
                                            apply Set.mem_of_mem_of_subset
                                            assumption'
        }
\end{lstlisting}
and prove that the function given by:\\
\begin{lstlisting}
    def Vh (φ : Formula) (F : @prime_filters α _) (I : Var → α) : Prop :=
        AlgInterpretation I φ ∈ F.1
\end{lstlisting}
is a valuation function for this frame.\\
Now, we can state and prove the second relation between algebraic and Kripke models:\\
\begin{lstlisting}
    lemma alg_kripke (I : Var → α) (φ : Formula) :
      true_in_alg_model I φ ↔ valid_in_model (prime_filters_frame I) φ
\end{lstlisting}
Finally, having this auxiliary results at hand, we can immediately prove the
equivalence between Kripke and algebraic validity:\\
\begin{lstlisting}
    theorem alg_kripke_valid_equiv (φ : Formula) :
      alg_valid φ ↔ valid φ :=
        by
          apply Iff.intro
          · intro Halg _ _
            rw [kripke_alg]; apply Halg
          · intro Hvalid _ _ _
\end{lstlisting}
\begin{lstlisting}
            rw [alg_kripke]; apply Hvalid
\end{lstlisting}

\section{Conclusion and future work}
We have used the Lean proof assistant to formally verify the completeness of IPL. After defining
the language, we formalized the Hilbert-style proof system and used it to establish a collection of
syntactic theorems and derived deduction rules. The next crucial step was formally specifying the two
studied semantics: Kripke and algebraic. For the proof of the completeness theorem with respect to
the Kripke semantics, we defined the so-called canonical model, and used it in order to complete the
proof by contraposition. On the other hand, for the algebraic completeness proof, we made use of the
Lindenbaum- Tarski algebra and some of its specific properties.

As future work, we aim to extend the current formalization to express Intuitionistic First-Order
Logic and also provide a completeness proof for this more complex system.
Furthermore, we intend to implement in Lean formal systems for intuitionistic arithmetical analysis and associated proof interpretations, as the ones presented in \cite{troelstra}.
\section{Acknowledgements}
The author thanks Laurențiu Leuștean and Traian Șerbănuță for providing comments and suggestions that improved the final version of the paper.

\bibliographystyle{eptcs}
\bibliography{bibliography}
\end{document}